\documentclass[12pt]{iopart}
\usepackage[numbers]{natbib}

\usepackage[dvips]{epsfig}
\usepackage[dvips,usenames]{color}
\usepackage{float}

\newcommand{\xpa}{x_\parallel}
\newcommand{\xpe}{x_\perp}

\begin{document}

\title{Percolation analysis of force networks in anisotropic granular
  matter}

\author{Romualdo Pastor-Satorras$^1$ and M.-Carmen Miguel$^2$}

\address{$^1$ Departament de F\'\i sica i Enginyeria Nuclear, Universitat
  Polit\`ecnica de Catalunya, Campus Nord B4, 08034 Barcelona, Spain}
  \address{$^2$ Departament de F\'\i sica
  Fonamental, Facultat de F\'\i sica, Universitat de Barcelona, Av.
  Diagonal 645, 08028 Barcelona, Spain}

\begin{abstract}
  We study the percolation properties of force networks in an
  anisotropic model for granular packings, the so-called
  q-model. Following the original recipe of Ostojic {\em et al.}
  [Nature \textbf{439} 828 (2006)], we consider a percolation process
  in which forces smaller than a given threshold $f$ are deleted in
  the network. For a critical threshold $f_c$, the system experiences
  a transition akin to percolation. We determine the point of this
  transition and its characteristic critical exponents applying a
  finite-size scaling analysis that takes explicitly into account the
  directed nature of the q-model. By means of extensive numerical
  simulations, we show that this percolation transition is strongly
  affected by the anisotropic nature of the model, yielding
  characteristic exponents which are neither those found in isotropic
  granular systems nor those in the directed version of standard
  percolation. The differences shown by the computed exponents can be
  related to the presence of strong directed correlations and mass
  conservation laws in the model under scrutiny.
  \end{abstract}


\maketitle

\section{Introduction}

Granular media show a peculiar kinetic behavior including the
possibility of exhibiting jammed configurations. Jammed assemblies of
grains at high densities are not able to explore the phase space but
can eventually yield at high drives, for instance under shear stress,
like a viscoplastic solid or a complex fluid
\cite{Liu01,Sitges19}. Over the last years, experimental observations
and numerical simulations of jammed granular media have repeatedly
shown the heterogeneous distribution of stress and contact forces in
dense packings \cite{Jaeger99}. Starting from the first studies of
weight distributions in bead packs \cite{oldpapers1,oldpapers2}, the
presence of force chains has been especially emphasized, chains which
form an intricate force network structure and are responsible for most
of the material's unusual properties. Force networks in a dense
granular packing play the role of the cytoskeleton in a living cell,
thus determining its mechanical response and stability. They
are also at the core of several important properties of granular media
such as friction and wear \cite{Marone98}, sound transmission
\cite{liu93}, or even electrical transport \cite{Wandewalle01}.

Internal stress and contact forces can be determined experimentally,
using for example photo-elastic materials, which exhibit
stress-induced birefringence.  The results obtained from birefringent
packings confirm that large forces seem to indeed concentrate along
branching-like paths, i.e. force chains or arches. Following some of
these measurements, it was argued that a close inspection of contact
force properties (for instance, the shape of force probability
distributions) could provide new insights regarding the
jamming-yielding transition in granular matter
\cite{Corwin05,Majmudar05}.  Nevertheless, the distribution of forces
alone does not describe the rich topological features observed in
experiments nor their potential physical consequences, and
complementary methods are thus required for their analysis.

The force network in a granular system is usually defined by the
contacts exerted between pairs of particles in the bulk of the system,
in such a way that, if particles $i$ and $j$ are in contact, they
mutually exert a symmetric force $f_{ij}$ that can also include
elastic and/or friction interactions. We can represent these pairwise
interactions in terms of a graph or network \cite{Newman2010}, in
which vertices represent the particles, and two vertices are joined by
an edge if the respective particles are in contact. This force network
can be further characterized as a weighted network, in which each edge
has assigned a real value $f_{ij}$, representing the actual value of
the force exerted by the vertices (the particles) $i$ and $j$ at the
ends of the edge.

Recently, Ostojic \textit{et al.} \cite{ostojic06,ostojic05} proposed
a novel way to obtain information about the structure of force
networks in static granular matter. The method is formulated in analogy
with percolation theory \cite{havlinpercolation,stauffer94} and is
based on the scaling properties of clusters of particles connected by
relatively large forces.  Since each edge carries a force $f_{ij}$, a
natural way to visualize the paths that carry the largest weight
(arches) is to consider only those edges with a force larger than a
given threshold $f$, $f_{ij} > f$, deleting those with $f_{ij} <
f$. For small values of $f$, essentially all forces remain in the
system, and they form a connected network with a single cluster
encompassing all the particles in the system.  Upon increasing the
value of $f$, the network is expected to break down in subnetworks of
connected forces, each representing a path of large weight. Each one
of these subnetworks can be understood in terms of clusters in a
percolation problem \cite{havlinpercolation}. By analogy with the
standard percolation transition, one expects to find a critical
percolation threshold $f_c$, such that for $f > f_c$ the force network
is fragmented into a large number of small clusters, while for $f =
f_c$ a large spanning cluster develops, reaching the boundaries of the
system.  This analogy with a percolation transition makes it possible
to characterize complex contact force networks in terms of a reduced
number of critical exponents \cite{stauffer94}.

In Ref.~\cite{ostojic06} the percolation transition in
contact force networks was first studied by applying a finite-size
scaling (FSS) \cite{privman90} data collapse technique. This
technique allows to estimate the value of the percolation threshold
$f_c$ as well as some exponents related to the divergence of the
average cluster size in the infinite network size limit. The
remarkable conclusion of this work is that different isotropic
models of a dense granular packing seem to exhibit similar percolation
exponent values, independently of their microscopic details.  Thus
these exponents appear to define a robust new universality class for
contact force networks, a class which, on the other hand, is different
from that of standard percolation. 

Many real granular systems, however, are strongly anisotropic; for
example, sand piles and silos are driven by the action of gravity, and
have therefore a preferred (downwards) direction. The presence of
anisotropy should in these other cases be naturally reflected in the
contact force network percolation transition, making it in principle
more akin to the anisotropic counterpart of percolation, namely
directed percolation \cite{kinzel83}. In fact, in
Ref.~\cite{ostojic05} (see also \cite{ostojic06}) it was observed that
anisotropic packing models indeed exhibit a different scaling in their
force network percolation transition\footnote{On the contrary,
  Ref.~\cite{Ostojic_shear} considered granular packings under the
  anisotropic effects induced by the application of a shear stress,
  concluding that shear-induced anisotropy was not enough to modify
  the universal exponents observed in the isotropic case.}. The
results in Refs.~\cite{ostojic06,ostojic05} however, were based in the
application of an intrinsically isotropic formalism to an anisotropic
system, not taking into account, for example, that correlation lengths
along different directions might scale differently.

Our purpose in this paper is to fill in this gap, presenting a
detailed study of the force network percolation transition in an
anisotropic system, performing a direct anisotropic scaling analysis.
We focus on the q-model \cite{liuoriginal}, a toy granular model
intended to represent the behavior of silos, having a clearly defined
preferred direction, in which the weight of the particles is
transmitted by virtue of gravity. Performing a detailed FSS numerical
analysis we uncover the anisotropic nature of this model, which shows
up mainly in the presence of two correlation lengths, with different
scaling behavior near the percolation threshold. Our numerical
simulations allow us to determine a number of critical exponents,
which we compare with those of directed percolation. The quantitative
differences observed in the exponents clearly indicate that the
contact force network percolation transition in granular systems with
a preferred direction belongs to a new anisotropic universality class,
which we fully characterize in terms of its critical exponents.

The present paper is organized as follows: In Sec.~\ref{sec:q-model}
we briefly review the definition of the q-model used in our study.
Section~\ref{sec:anis-perc-analys} describes the main elements of the
FSS theory for anisotropic systems. The results of our analysis are
presented in Section~\ref{sec:computer-simulations}.  Finally, in
Sec.~\ref{sec:disc-concl} we summarize our results and present our
conclusions and perspectives.

\section{The q-model}
\label{sec:q-model}

The q-model \cite{liuoriginal} is defined on a tilted two-dimensional
square lattice, whose sites are labeled by two integer numbers,
$(\xpa, \xpe)$, $\xpa=1,\ldots L_\parallel$, $\xpe=1,\ldots L_\perp$,
giving its vertical and horizontal position, respectively.  Each site
in the row $\xpa$ supports the weight of its two nearest neighbors in
the immediate upper row $\xpa-1$. Simultaneously, its own total weight
is distributed between its two nearest downward neighbors located in
row $\xpa+1$.  The transmission of weight from one row to the next is
thus given by the equation
\begin{equation}
  w(\xpa, \xpe) = w_0 + P \delta_{\xpa,1}+
  \sum_{\alpha=-1}^{+1}q_\alpha(\xpa-1, 
  \xpe-\alpha) w(\xpa-1,\xpe-\alpha)  
\label{eq:10}
\end{equation}
where $w_0$ is the constant weight contributed by each single site,
$P$ is a constant pressure downwards applied at the topmost row, and
$q_\alpha(\xpa,\xpe)$, with $\alpha=\pm1$, are uniformly distributed
random numbers between zero and one, restricted by the mass
conservation condition $\sum_{\alpha}q_\alpha(\xpa,\xpe)=1$.
Eq.~(\ref{eq:10}) determines the set of weights $w(\xpa, \xpe)$
corresponding to an equilibrium configuration, as well as the
corresponding force network. For instance, the relative forces between
a particle at $(\xpa, \xpe)$ and its upward neighbors
$(\xpa-1,\xpe-\alpha)$ are given by $q_\alpha(\xpa-1, \xpe-\alpha)
w(\xpa-1,\xpe-\alpha)$.

In the following we will consider the q-model defined on a lattice
with periodic boundary conditions along the $\xpe$ axis
\cite{nicodemi}, with massless particles $w_0=0$ and constant $P$. In
this case, a system of linear dimensions $L_\parallel$ and $L_\perp$
contains $L_\parallel L_\perp /2$ particles, each row bears an
average constant weight per particle $P$, i.e.  no weight is lost at
the system boundaries, and the average force between particles is
$\langle f_{ij}\rangle = P/2$. Obviously, the pressure $P$ is just a
rescaling factor in all forces, so we set it equal to one, without
loss of generality.

\section{Anisotropic finite-size scaling analysis}
\label{sec:anis-perc-analys}

In this section we review the FSS theory needed
to analyze the force network percolation transition in an anisotropic
system such as the scalar q-model.  Let is first consider the
isotropic case, in which there is a single correlation length $\xi$,
diverging as $\xi \sim \Delta^{-\nu}$ as a function of the distance to
the percolation threshold $\Delta=|f-f_c|$. Information about the
position of the critical point and exponent values can be obtained by
studying the normalized cluster number $n(s, f)$, defined as the
number of clusters of size $s$ per lattice site \cite{stauffer94}.
For this purpose, we define the average cluster size (or
susceptibility)
\begin{equation}
  \chi(f) = \sum_s  s^2 n(s, f).
  \label{eq:4}
\end{equation}
In an infinite system, and close to the percolation threshold, the
susceptibility diverges as $\chi(f) \sim \Delta^{-\gamma}$. In a
finite system of length $L$, the FSS hypothesis \cite{privman90}
states that the only relevant length scale is $\xi$, and that the
system size dependence can only enter through the ratio $\xi/
L$. Thus, at finite $L$ the susceptibility scales as
\begin{equation}
   \chi(f,L) = L^{\gamma / \nu} \chi_0(\Delta^{\nu} L),
   \label{eq:5}
\end{equation}
where $ \chi_0(x) \to x^{-\gamma/ \nu}$ for $x\to\infty$, and $
\chi_0(x) \to \mathrm{const.}$ for $x\to0$.  Thus, for $\Delta=0$,
$\chi(f_c,L)$ would grow as a pure power law with $L$, while for
$\Delta\neq0$ it would deviate from the power law behavior and
saturate to a constant value for sufficiently large $L$. An estimate
of $f_c$ can be obtained as the the one yielding the best power law
fit to $\chi(f,L)$ as a function of $L$. Once $f_c$ is determined, a
linear regression provides an estimate of the exponent ratio $\gamma /
\nu$. Additional exponents (and exponent relations) can be computed
from a closer examination of the normalized cluster number. In fact,
close to the percolation threshold, the normalized cluster number
scales as \cite{stauffer94}
\begin{equation}
  n(s, f) = s^{-\tau} \mathcal{F}(s \Delta^{1/ \sigma}),
\end{equation}
where $\sigma$ is a critical exponent giving the characteristic
cluster size, $s_c \sim \Delta^{-1/\sigma}$, and $\mathcal{F}$ is a
universal function, independent of $s$ and $\Delta$. Substituting
$\Delta \sim \xi^{-1/\nu}$, and defining the fractal dimension $D$ as
$s_c \sim \xi^D$, one obtains $ D=1/(\sigma \nu)$.  Right at the percolation
threshold, in a system of finite size $L$, the cluster number will
scale as
\begin{equation}
  n(s, f_c, L) = s^{-\tau} f(s L^{-D}),
\end{equation}
and, from Eq.~(\ref{eq:4}), and comparing with Eq.~(\ref{eq:5}), we
obtain the  scaling relation
\begin{equation}
  \gamma = \frac{3-\tau}{\sigma} = (3-\tau)D \nu.
  \label{eq:6}
\end{equation}

The critical point and some critical exponents can also be estimated
by means of a bisection method
\cite{springerlink:10.1007/BFb0012540,williams84:_direc}. Consider a
random realization of a force network with size $L$ and an initial
guess for the percolation threshold $f_c^{0}=f_{max}/2$, where
$f_{max}$ is the maximum force present in the network. We can estimate
the true percolation threshold by an iterative procedure. In any step
with a guess value $f_c^{i}$, we check whether a percolating
(spanning) cluster exists or not. If it does, we increase the
threshold by $f_c^{i+1}=f_c^{i}+f_{max}2^{-(i+1)}$; otherwise we
decrease it as $f_c^{i+1}=f_c^{i}-f_{max} 2^{-(i+1)}$. Iterating this
scheme a sufficient number of times, we compute the percolation
threshold for a given network realization. Averaging over many random
networks, we can obtain an estimate of the threshold $\langle
f_c(L)\rangle$ for the system size considered. The fluctuations of
this estimate, $\sigma(L) = \left[\langle f_c(L)^2 \rangle - \langle
  f_c(L)\rangle^2\right]^{1/2}$, as a function of $L$, yields the
value of the correlation exponent,
\begin{equation}
  \sigma(L) \sim L^{-1/ \nu},
\end{equation}
while the percolation threshold can be obtained from the average value
as
\begin{equation}
  | f_c - \langle f_c(L)\rangle| \sim  L^{-1/ \nu}.
\end{equation}

In anisotropic systems, the FSS theory takes a slightly more complex
form. The length of a typical cluster is now given by the correlation
lengths along the longitudinal (downwards) and transverse directions,
$\xi_\parallel$ and $\xi_\perp$, respectively, that scale as
\begin{equation}
  \xi_\parallel \sim \Delta^{-\nu_\parallel}, \qquad \mathrm{and}
  \qquad \xi_\perp \sim \Delta^{-\nu_\perp},
  \label{eq:3}
\end{equation}
where the exponents $\nu_\perp$ and $\nu_\parallel$ are, in principle,
different. The anisotropy exponent, measuring the differente scaling
of both correlation lengths, is defined as the ratio
\begin{equation}
  \theta =\frac{\nu_\parallel}{\nu_\perp}. 
  \label{eq:2}
\end{equation}
In finite size simulations, two different length scales are thus
present, $L_\parallel$ and $L_\perp$. Varying them independently would
lead to an uncontrolled scaling of the relevant functions. 
A proper 
analysis \cite{binder89:_finit} shows, however, that when the longitudinal and
perpendicular lengths are related by the constraint
\cite{williams84:_direc,redner82:_conduc,wang96:_anisot}
\begin{equation}
  L_\parallel \sim L_\perp^\theta,
  \label{eq:11}
\end{equation}
the system behaves as if effectively isotropic, and standard FSS
applies in terms of a single length scale.  This fact suggest an
efficient way to compute the critical percolation exponents by
performing numerical simulations for systems with freely varing $
L_\parallel$, and fixed $L_\perp = L_\parallel^{1/\theta}$.  With now
a single characteristic length, the percolation threshold and the
exponent ratio $\gamma/ \nu_\parallel$ can be found by a standard FSS
analysis of the susceptibility, which at the critical point takes the
form \cite{binder89:_finit}
\begin{equation}
  \chi(f_c, L_\parallel,L_\parallel^{1/
  \theta}) \sim L_\parallel^{\gamma/\nu_\parallel}.
\end{equation}
Analogously, the normalized cluster number will take the form
\begin{equation}
  n(s, f_c,  L_\parallel,L_\parallel^{1/
    \theta})) = s^{-\tau} f(s L_\parallel^{-D_\parallel}),
  \label{eq:9}
\end{equation}
where the exponent $D_\parallel$ will satisfy the anisotropic
equivalent of Eq.~(\ref{eq:6}), namely
\begin{equation}
  \gamma= (3-\tau) D_\parallel \nu_\parallel.
\end{equation}

The bisection method described above can also be analogously modified
to work in anisotropic systems~\cite{williams84:_direc}. With the
rescaling of system lengths given by Eq.~(\ref{eq:11}) the variance of
the threshold estimate at finite sizes takes the form
\begin{equation}
  \sigma(L_\parallel,L_\parallel^{1/ \theta}) = \left[\langle
  f_c(L_\parallel,L_\parallel^{1/ \theta})^2 \rangle -  \langle
  f_c(L_\parallel,L_\parallel^{1/ 
    \theta}) \rangle^2 \right]^{1/2}\sim  L_\parallel^{-1/ \nu_\parallel},
\end{equation}
and the percolation threshold is given by
\begin{equation}
  | f_c -  \langle f_c(L_\parallel,L_\parallel^{1/ \theta})\rangle|
  \sim   L_\parallel^{-1/ \nu_\parallel}. 
\end{equation}

\section{Computer simulations}
\label{sec:computer-simulations}

We have studied the percolation transition in the force network of the
scalar q-model by means of computer simulations on systems of size
$L_\parallel \times L_\perp$, with $L_\parallel$ up to $66451$ and
$L_\perp$ up to $2048$. In order to apply the anisotropic FSS scheme
described above, they key point is to have an \textit{a priori}
knowledge of the anisotropy exponent $\theta$. A numerical estimate of
this exponent can be obtained using the fact that, close to the
critical point, the two correlation lengths must be related by
$\xi_\parallel \sim \xi_\perp^\theta$. Consider a system of very large
longitudinal size $L_\parallel$ and a small transversal size $L_\perp
\ll L_\parallel$. In the vicinity of the percolation threshold, for
small $L_\perp$ we will have $\xi_\perp \sim L_\perp$, and by
increasing $L_\perp$, we will observe that $\xi_\parallel$ increases
as $\xi_\parallel \sim L_\perp^\theta \sim \xi_\perp^\theta$. For
sufficiently large $L_\perp$, and not too close to the threshold, we
will have that both $\xi_\parallel$ and $\xi_\perp$ saturate to their
corresponding values given by Eq. (\ref{eq:3}).  Therefore, the
exponent $\theta$ can be determined by simulations at fixed and large
$L_\parallel$, by plotting $\xi_\parallel$ as a function of
$\xi_\perp$ computed for increasing, but small, $L_\perp$ values, and
different force thresholds. The $f$ values yielding the best power law
fits are in the vicinity of the percolation threshold $f_c$.

\begin{figure}[t]
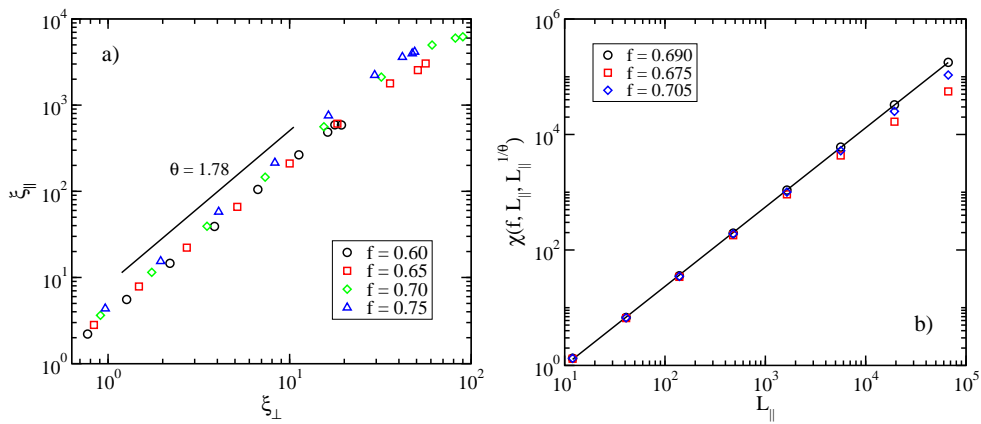

  \centerline{\epsfig{file=Fig1a.eps, height=5.5cm}
    \epsfig{file=Fig1b.eps, height=5.5cm}}
  \caption{a) Correlation lengths in the q-model, computed at fixed
    $L_\parallel = 16384$ and variable $L_\perp \ll L_\parallel$, for
    different values of the threshold force $f$. b) Susceptibility
    $\chi(f, L_\parallel,L_\parallel^{1/ \theta})$ of the q-model for
    different values of $f$. The straight line corresponds to the best
    power-law fit, corresponding to $f=0.690$ and yielding a slope
    $\gamma / \nu_\parallel \simeq 1.37$.}
   \label{fig:Correl}
\end{figure}

In Fig.~\ref{fig:Correl}(a), we present the results of simulations of the
q-model with fixed $L_\parallel = 16384$ and $L_\perp$ running from
$16$ up to $2048$, for different values of $f$. Correlation lengths
were computed as is customarily done in anisotropic systems
\cite{rvdirected}: For each cluster $c$ of connected forces
that is composed by a set of vertices $\{x_\parallel^{(i)},
x_\perp^{(i)}\}$, with $i=1,\ldots,s$, we define the quantities
\begin{equation}
  R_\parallel(c) = \frac{1}{s} \sum_{i=1}^s |x_\parallel^m -
  x_\parallel^{(i)}|, \qquad
  R_\perp^2(c) = \frac{1}{s} \sum_{i=1}^s (x_\perp^m -
  x_\perp^{(i)})^2, 
\end{equation}
where $x_\parallel^m$ and $x_\perp^m$ are the coordinates of some
 reference point within the cluster. We have chosen the point
with the highest longitudinal coordinate and the average $x_\perp$
coordinate, respectively. Then, the correlation lengths are defined as
\begin{equation}
  \xi_\parallel = \frac{\sum_s^{\prime} R_\parallel(s) s^2 n(s,
    f)}{\sum_s^{\prime} s^2 
    n(s,f)}, \qquad   \xi_\perp^2 = 
  \frac{\sum_s^{\prime} R_\perp^2(s) s^2 n(s, f)}{\sum_s^{\prime} s^2
    n(s,f)}, 
\end{equation}
where the prime indicates that one has to exclude the spanning
clusters from the sum over cluster sizes. From the plots in
Fig. \ref{fig:Correl}(a), in which we have represented the data
providing a best power-law fitting, we conclude that the percolation
threshold is located in the vicinity of $f\sim 0.70$. Moreover, a
linear regression for the smallest values of $\xi_\perp$ yields an
estimate of the anisotropy exponent $\theta = 1.78 \pm 0.05$.

Once the exponent $\theta$ has been estimated, we can proceed with the
full FSS analysis. In the first place, we focus on the behavior of the
susceptibility $\chi(f, L_\parallel,L_\parallel^{1/ \theta})$ computed
for $L_\parallel$ ranging from $12$ to $66451$.  In
Fig.~\ref{fig:Correl}(b) we represent the susceptibility as a function
of $L_\parallel$ for different values of $f$. As can be seen in the
plot, the best power law behavior for $\chi(f,
L_\parallel,L_\parallel^{1/ \theta})$ is obtained for the threshold
force $f_c=0.690\pm0.005$; significant deviations can be observed for
slightly larger and smaller values of $f$. A linear regression of
$\chi(f, L_\parallel,L_\parallel^{1/ \theta})$ at the percolation
threshold yields the exponent ratio $\gamma / \nu_\parallel = 1.37 \pm
0.01$.

The numerical analysis of the full normalized cluster size distribution at
the percolation threshold can be performed using the moment analysis
technique developed for the study of self-organized critical systems
\cite{men}.  The $k$-th moment $M_k$ of the cluster distribution is
defined as
\begin{equation}
  M_k = \sum_s s^k  n(s, f_c,  L_\parallel,L_\parallel^{1/
    \theta})),
  \label{eq:8}
\end{equation}
At the percolation threshold, when the cluster number is given by
Eq.~(\ref{eq:9}), we have that $M_k(L_\parallel) \sim
L_\parallel^{\alpha(k)}$, where the $k$-dependent exponent is given by
\begin{equation}
  \alpha(k) = D_\parallel k + D_\parallel(1-\tau).
  \label{eq:12}
\end{equation}
Thus, computing $M_k(L_\parallel)$ as a function of $L_\parallel$ for
different system sizes provides information on $\alpha(k)$, which
should be a linear function of $k$ of the form $\alpha(k) = \alpha_0 +
k \alpha_1$, from which we obtain $D_\parallel = \alpha_1$ and $\tau =
1-\alpha_0 / \alpha_1$. The correctedness of exponent's values can be
checked by means of a data collapse technique: Noticing that the
normalized cluster number $n(s, f_c)$ scales as given by
Eq.(\ref{eq:9}), then $L_\parallel^{\tau D_\parallel} n(s, f_c)$
should collapse onto a universal function when plotted as a function
of the rescaled variable $s L_\parallel^{-D_\parallel}$.
\begin{figure}[t]
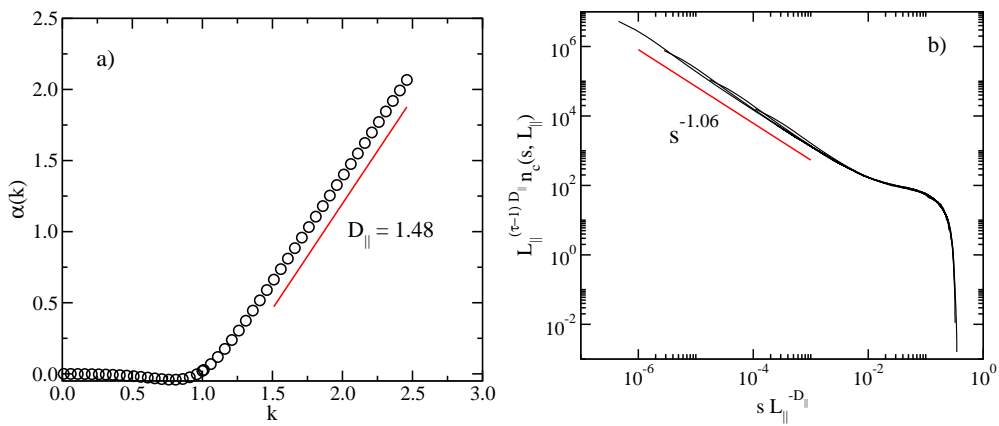

  \centerline{\epsfig{file=Fig2a.eps, height=5.5cm}
    \epsfig{file=Fig2b.eps, height=5.5cm}}
  \caption{a) Plot of the $\alpha(k)$ functions for the q-model at the
    percolation threshold. The straight line is a least-squares
    fitting yielding the corresponding $D_\parallel$. b) Data collapse
    analysis of the integrated cluster number for the q-model at the
    percolation threshold. Systems sizes are $L_\parallel=478$,
    $1641$, $5634$, and $19349$.}
   \label{fig:Alpha}
\end{figure}

In Fig.~\ref{fig:Alpha}(a) we plot the $\alpha(k)$ evaluated from
linear regressions of the moments $M_k(L_\parallel)$, computed from
numerical simulations at the percolation threshold with system sizes
$L_\parallel=139$, $478$, $1641$, $5634$, and $19349$. A linear
regression of this function, provides the values $D_\parallel =
\alpha_1 = 1.48 \pm 0.01$, and $\alpha_0 = -1.57\pm0.01$, from which
we obtain $\tau = 2.06 \pm 0.02$. This last value can be checked
against the scaling relation Eq.~(\ref{eq:6}) (properly redefined for
anisotropic systems), which leads to $\tau = 3 - \gamma/(D_\parallel
\nu_\parallel) \simeq 2.07$, in perfect agreement with the estimate
from the regression of the $\alpha(k)$ function.  In order to check
the accuracy of these exponents for the q-model we perform a data
collapse analysis of the integrated cluster number at the percolation
threshold, defined as
\begin{equation}
  n_c(s, L_\parallel) = \sum_{s'=s}^\infty n(s,L_\parallel).
\end{equation}
In Fig.\ref{fig:Alpha}(b) we observe that, as expected, the plots of
the integrated cluster number, under the rescaling $s \to s
L_\parallel^{-D_\parallel}$ and $n_c(s, L_\parallel) \to
L_\parallel^{(\tau -1)D_\parallel} n_c(s, L_\parallel)$, collapse onto
a single universal function for different values of $L_\parallel$.

\begin{figure}[t]
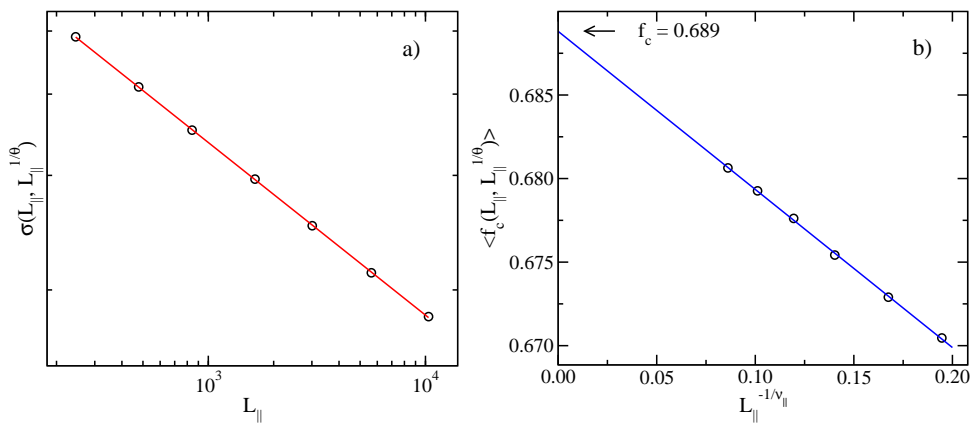

 \centerline{\epsfig{file=Fig3a.eps, height=5.5cm}
   \epsfig{file=Fig3b.eps, height=5.5cm}}
  \caption{a) Fluctuations of the percolation threshold estimated by means of
  the bisection method. 
  b) Extrapolation of the critical point from the bisection method.}
   \label{fig:Regression}
\end{figure}
We turn now our attention to the application of the bisection method.
Fig.~\ref{fig:Regression}(a) shows the fluctuations
$\sigma(L_\parallel,L_\parallel^{1/ \theta})$ computed as a function
of $L_\parallel$. A linear regression provides the slope $1/
\nu_\parallel$, from which we estimate the corresponding 
critical exponent $ \nu_\parallel = 3.77 \pm 0.01$. With this result,
we can compute the exponent $\gamma$ from the ratio $\gamma /
\nu_\parallel$, obtaining $\gamma = 5.18 \pm 0.04$, and from
Eq.~(\ref{eq:2}), $\nu_\perp = 2.12 \pm 0.01$. Finally, using the
previously computed exponent, we can plot $\langle
f_c(L_\parallel,L_\parallel^{1/ \theta})\rangle$ as a function of
$L_\parallel^{-1/\nu_\parallel}$, as in Fig.~\ref{fig:Regression}(b), which
shows a good linear behavior, with an intercept with the vertical axis
providing the value $f_c \simeq 0.689$ in excellent agreement with the
threshold obtained from the analysis of the susceptibility.

\section{Summary and discussion}
\label{sec:disc-concl}

In Table~\ref{tab:exponents} we summarize the results we have obtained
in our percolation analysis of the contact force network in the
anisotropic q-model, compared with the exponents for isotropic and
directed percolation, and with the exponents (or exponent ratios)
available for the percolation transition in isotropic contact force
networks \cite{ostojic06,ostojic05}.
\begin{table}[t]
  \caption{Critical exponents for the percolation transition in the
    contact force network of q-model, compared with the
    values corresponding to 
    isotropic percolation (IP) in two dimensions,
    directed percolation (DP) percolation in 1+1  dimensions, and the
    percolation transition in isotropic contact force networks
    (ICFN). Exponents from 
    Refs.~\cite{stauffer94,munoz99:_avalan,ostojic06}.} 
  \label{tab:exponents}
  \begin{center}
    \begin{tabular}{c||c||c|c||c}
      Exponent &  q-model  & IP  &  DP & ICFN\\  \hline \hline
      $\gamma$ &  $5.18 \pm 0.04$ & $43/18 = 2.3889$  & $0.54386$ &
      $2.8\pm0.2$\\ 
      $\nu_\parallel$  &   $3.77 \pm 0.01$  & $4/3= 1.3333$ & 
      $1.73383$ & $1.6\pm0.1$\\
      $\nu_\perp$  &   $2.12 \pm 0.01$  & $4/3=1.3333$ & $1.09684$ & $1.6\pm0.1$\\
      $D_\parallel$  & $1.48 \pm 0.01$   & $91/48= 1.8958$&
      $1.4727$ & --- \\
      $\tau$  &  $2.06 \pm 0.02$  & $187/91= 2.0549$ & $2.108$ & --- \\ \hline
      $\gamma/\nu_\parallel$  &  $1.37\pm0.01$  & $43/24 = 1.7917$ &
      $0.3137$ & $1.78\pm0.02$\\ 
    \end{tabular}
  \end{center}
\end{table}
We note that the results obtained here for the q-model are compatible
with those reported in Refs.\cite{ostojic06,ostojic05}, namely $f_c
\simeq 0.70$, $\gamma/\nu \simeq 1.38$, and $\nu \simeq 3.1$. Our
method for estimating exponents is however more accurate and
systematic, being at the same time capable of providing new exponents,
not considered previously. This is specially evident for the exponent
$\nu=3.1\pm0.1$ quoted in \cite{ostojic06}, which is not discerning
between the parallel and perpendicular directions.

The main conclusion extracted from the analysis of these exponents is
that, at least in two dimensions, the percolation transition in the
contact force network of anisotropic granular matter belongs to a
universality class different from either anisotropic contact force
networks and isotropic percolation. It is noteworthy that the change
of universality goes thus beyond the simple presence of a preferred
direction, as we can see from the comparison of the q-model exponents
with those of directed percolation. Even though some exponents are
similar, such as $\tau$ or $D_\parallel$, others are clearly
different, out of the estimated error bars. The ultimate reason for
this difference can be traced out in the presence of force
correlations or arches \cite{nicodemi}.  The strength of these arches
is enhanced in anisotropic models with a preferred direction for the
propagation of weight, and explain the change in universality between
different packing models.  The origin of correlations is easy to
understand in the present case: As we have defined it, the total force
between rows is constant, imposing a global conservation law,
superimposed to the local conservation of weights built in the
definition of the model, Eq.~(\ref{eq:10}). Global conservation
prevents dissipation of stresses, and as a consequence any local build
up of forces will propagate downwards unchecked and lead to the
creation of arches in which strong forces are preferably connected to
one another.
\begin{figure}[t]
  \centerline{\epsfig{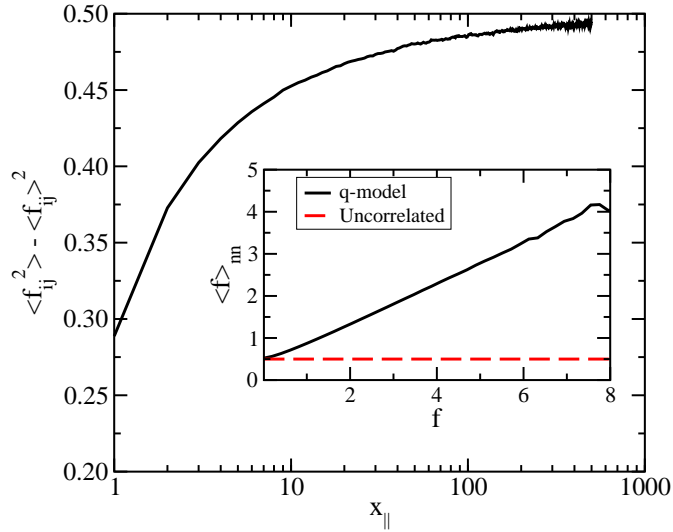}}
  \caption{ Standard deviation of the force distribution as a function
    of the longitudinal coordinate $\xpa$ article forces for two
    values of the longitudinal direction $\xpa$ in the massless
    q-model of size $500 \times 500$, averaged over $500$ system
    realizations. Pressure per particle $P=1$.. Inset: Average nearest
    neighbor forces as a function of the force $f$.}
   \label{fig:Fdist}
\end{figure}
The strong anisotropy and correlations in the force network of the
q-model are checked in Fig.~\ref{fig:Fdist}, where we plot the
variance of the force distribution, computed at different heights
$\xpa$, and which shows a marked dependence on this variable. In the
case forces were uncorrelated at different $\xpa$ levels, and
considering that forces are exponentially distributed
\cite{ostojic06}, the variance should take the form $P^2/4 = 0.25$,
clearly smaller than the numerically computed values. On the other
hand, correlations between nearest neighbors are checked in the inset
of Fig.~\ref{fig:Fdist}, where we plot the average value $\langle
f\rangle_{nn}$ of the forces connected to a given bond of force $f$
\cite{alexei}.  As we can see, this average value grows almost
linearly with $f$, while in absence of correlations it should be equal
to the average force $\langle f_{ij}\rangle = P/2 =0.5$.

We conclude therefore that force networks in granular matter define
different universality classes, depending on the symmetries imposed on
the systems, universality classes that bear no resemblance with the
corresponding ones in standard percolation, and are strongly affected
by the strength of  correlations in the overall force network
structure. This result calls for further research in order to clarify
the situation in more realistic settings, where the anisotropy might
not be as strong as in the simple q-model \cite{Ostojic_shear}.

\section*{Acknowledgements}
R. P.-S. acknowledges financial support from the Spanish MEC, under
project No. FIS2010-21781-C02-01, as well as additional support
through ICREA Academia, funded by the Generalitat de
Catalunya. M.-C. M. acknowledges financial support from the Spanish
MEC, under project No. FIS2010-21781-C02-02, as well as additional
support through the I3 program.



\providecommand{\newblock}{}

\end{document}